\documentclass[a4paper, twocolumn, prb]{revtex4-2} 
\usepackage{color}
\usepackage{graphics}
\usepackage{epsfig}
\begin{document}
	
\title{Field induced spin freezing and low temperature heat capacity of disordered pyrochlore oxide Ho$_{2}$Zr$_{2}$O$_{7}$}
	
\author{Sheetal$^{1}$, A. Elghandour$^{2}$, R Klingeler$^{2}$ and C. S. Yadav$^{1*}$}
\affiliation{$^{1}$School of Basic Sciences, Indian Institute of Technology Mandi, Mandi-175075 (H.P.), India}
\affiliation{$^{2}$Kirchhoff Institute of Physics, Heidelberg University, INF 227, D-69120 Heidelberg, Germany}
	
\begin{abstract}
		
Spin ice materials are the model systems that have a zero-point entropy as \textit{T} $\rightarrow$ 0 K, owing to the frozen disordered states. Here, we chemically alter the well-known spin ice Ho$_{2}$Ti$_{2}$O$_{7}$ by replacing Ti sites with isovalent but larger Zr ion. Unlike the Ho$_{2}$Ti$_{2}$O$_{7}$ which is a pyrochlore material, Ho$_{2}$Zr$_{2}$O$_{7}$ crystallizes in disordered pyrochlore structure. We have performed detailed structural, ac magnetic susceptibility and heat capacity studies on Ho$_{2}$Zr$_{2}$O$_{7}$ to investigate the interplay of structural disorder and frustrated interactions. The zero-field ground state exhibits large magnetic susceptibility and remains dynamic down to 30 mK without showing Pauling's residual entropy. The dynamic state is suppressed continuously with the magnetic field and freezing transition evolves ($\sim$ 10 K) at a field of $\sim$ 10 kOe. These results suggest that the alteration of chemical order and local strain in Ho$_{2}$Ti$_{2}$O$_{7}$ prevents the development of spin ice state and provides a new material to study the geometrical frustration based on the structure.   
\end{abstract}
	
\maketitle
	
\section{Introduction}
Beginning with the identification of unusual disordered ground state in magnetic materials, frustration becomes a source of fascination among the scientists. In these systems, pyrochlore A$_{2}$Ti$_{2}$O$_{7}$ (A = Dy, Ho) as an exemplar of three-dimensional frustration continues to be the central topic of research due to the observation of water ice-like highly degenerate macroscopic ground state \cite{matsuhira2001novel,ramirez1999zero}. Dy$_{2}$Ti$_{2}$O$_{7}$ and Ho$_{2}$Ti$_{2}$O$_{7}$ are the well-known spin ice materials that possess the finite residual entropy at the lowest temperature which is equivalent to the Pauling's value of water ice \cite{ramirez1999zero,bramwell2001spin}. Apart from the spin-ice behavior below $\sim$ 2 K, Dy$_{2}$Ti$_{2}$O$_{7}$ also shows a strong frequency-dependent spin-freezing at $\sim$ 16 K \cite{snyder2001spin,snyder2004low}. However no such feature is seen for the Ho$_{2}$Ti$_{2}$O$_{7}$ which distinguishes the spin dynamics of (Ho/Dy)$_{2}$Ti$_{2}$O$_{7}$ \cite{lau2006zero}.\\
	
In order to understand the spin-ice state, researchers have begun to look outside the present structural map scenario by forming the stuffed pyrochlores by placing a non-magnetic atom or the atoms with smaller moments in the pyrochlore lattice \cite{kimura2013quantum,wen2017disordered,petit2016observation,ramon2019absence,sibille2017coulomb,anand2018optimization,anand2015investigations}. Various combinations of A and B site ions are employed to replace the existing spin ice state and disorder the symmetry of the available crystal structure. The zirconate and hafnates pyrochlore provides a new avenue to look for a similar magnetic ground state. The stability of the pyrochlore structure can be empirically investigated in terms of atomic radii ratio (\textit{r}$_{A}$/\textit{r}$_{B}$) of the respective A and B ions \cite{mouta2013tolerance}. The stable pyrochlore structure can be realized only for the range of 1.48 $\leq$ \textit{r}$_{A}$/\textit{r}$_{B}$ $\leq$ 1.71 ratio, and any deviation from this value disturbs the structural symmetry \cite{mouta2013tolerance, sheetal2020evolution}. For comparison, the radius ratio for Dy$_{2}$Zr$_{2}$O$_{7}$ and Ho$_{2}$Zr$_{2}$O$_{7}$ lies between 1.39 and 1.44, which is smaller than the nominal range for pyrochlore structure. These materials exhibit disordered pyrochlore structures and show the absence of low-temperature magnetism of rare-earth moments. For example, Pr$_{2}$Zr$_{2}$O$_{7}$ exhibits spin ice-like correlations and strong quantum fluctuations \cite{kimura2013quantum,wen2017disordered}. Nd$_{2}$Zr$_{2}$O$_{7}$ reveals the coexistence of an ordered and fluctuating Coulomb phase at low-temperature \cite{petit2016observation}. Dy$_{2}$Zr$_{2}$O$_{7}$, Tb$_{2}$Hf$_{2}$O$_{7}$ and Pr$_{2}$Hf$_{2}$O$_{7}$ remain dynamic down to 100 mk \cite{ramon2019absence,sibille2017coulomb,anand2018optimization}. However the Dy$_{2}$Zr$_{2}$O$_{7}$ and Pr$_{2}$Hf$_{2}$O$_{7}$ have been shown to exhibit glassy transition at $\sim$ 90 mK \cite{ramon2019absence,anand2016physical}. Nd$_{2}$Hf$_{2}$O$_{7}$ shows a long-range antiferromagnetic state ($T_{N}$ $\approx$ 0.55 K) with an all-in-all-out (AIAO) spin configuration \cite{anand2015observation}. Recent research also shows the loss of spin ice ground state due to a large chemical disorder induced by the stuffing of pyrochlores \cite{ramon2019absence,sheetal2020emergence}.\\

There are few reports on the evolution of structure and magnetism of the magnetically diluted pyrochlore lattice under the application of magnetic field \cite{xu2018field,sheetal2020evolution,zhang2008high}. It was found that the structural disorder and spin ice state can be stabilized in disordered pyrochlores either by the application of magnetic field or by suitable choice of non-magnetic substitution over A site \cite{sheetal2020emergence,sheetal2020evolution}. The experimental studies on Dy$_{2}$Zr$_{2}$O$_{7}$ and Nd$_{2}$Zr$_{2}$O$_{7}$ exhibit the field-induced behavior in quantum spin ice state and the emergent low-dimensional	magnetism \cite{sheetal2020emergence,xu2018field}. The substitution of non-magnetic La$^{3+}$ ion in Dy$_{2}$Zr$_{2}$O$_{7}$ induces the structural change from disordered pyrochlore to a stable pyrochlore structure and observation of spin freezing at \textit{H} = 0 Oe similar to the field-induced spin freezing of Dy$_{2}$Zr$_{2}$O$_{7}$/Ho$_{2}$Ti$_{2}$O$_{7}$ and the well-known spin ice system Dy$_{2}$Ti$_{2}$O$_{7}$ \cite{sheetal2020evolution,ehlers2004evidence}. In our previous work, we have reported the complete magnetic and structural phase diagram of  Dy$_{2}$Zr$_{2}$O$_{7}$ and the evolution of pyrochlore phase and spin freezing transition on diluting Dy$_{2}$Zr$_{2}$O$_{7}$ by non-magnetic La over Dy. Further, pressure effects exhibit weak pyrochlore-type ordering in case of Ho$_{2}$Zr$_{2}$O$_{7}$ and Er$_{2}$Zr$_{2}$O$_{7}$\cite{zhang2008high}.\\
	
In this paper, we have studied a new disordered pyrochlores zirconate Ho$_{2}$Zr$_{2}$O$_{7}$ and discussed its structural, magnetic, and thermodynamic properties. We observe that with the stuffing of Zr on the Ti site, the spin ice character is completely lost. The large values of susceptibility and heat capacity reveal a very dynamic ground state down to 30 mK. However, we found evidence of field-induced spin freezing at $\sim$ 10 K similar to the spin ice Ho$_{2}$Ti$_{2}$O$_{7}$. This also rules out the importance of high-temperature spin freezing in the formation of spin ice state below 1 K in Dy$_{2}$Ti$_{2}$O$_{7}$.
	
\section{Experimental Details}
	
The polycrystalline Ho$_{2}$Zr$_{2}$O$_{7}$ compound was prepared by solid-state reaction method by giving two heat treatments at 1350$^{o}$C with intermediate grindings and the final heat treatment to the pellets at the same temperature \cite{sheetal2020emergence}. X-ray diffraction (xrd) experiment was performed using Rigaku x-ray diffractometer in the 2$\theta$ range of 10-90$^{o}$. The crystal structure and phase purity were confirmed by performing Rietveld refinement of the powder sample using Fullprof Suit software and the graphical interface Vesta \cite{rodriguez1990fullprof}. The magnetic measurements were carried out using Quantum Design built Magnetic Property Measurement System (MPMS) down to 1.8 K. Heat capacity measurement was measured using Quantum Design built Physical Property Measurement System (PPMS) down to 30 mK.
	
\section{Results and Discussion}
	
	\begin{figure}[ht]
		\begin{center}
			\includegraphics[scale=0.35]{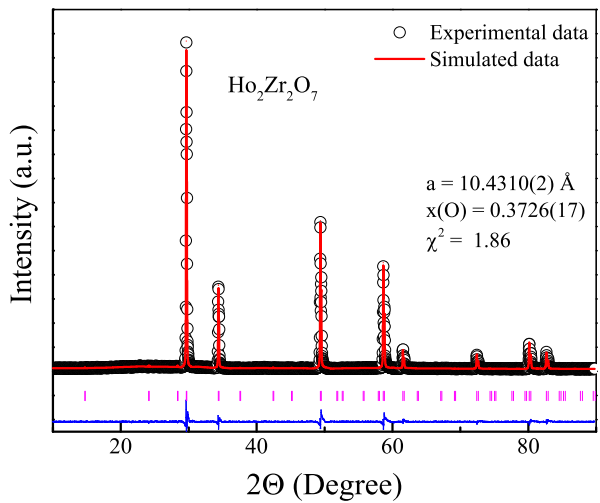}
			\caption{Room temperature x-ray powder diffraction pattern along with the Rietveld refined simulated data for Ho$_{2}$Zr$_{2}$O$_{7}$. As seen from the figure, the superstructure peaks corresponding to pyrochlore phase are missing.}
			\vspace{-15pt}
		\end{center}
	\end{figure}
	
	\begin{figure}[ht]
		\begin{center}
			\includegraphics[scale=0.4]{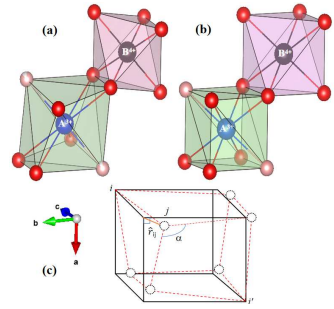}
			\caption{Coordination of A$^{3+}$ and B$^{4+}$ ions in A$_{2}$B$_{2}$O$_{7}$. (a) Pyrochlore structure: O$^{2-}$ ions around A$^{3+}$ (blue) rare-earth ion and B$^{4+}$ (black) transition-metal ion form the distorted cube and octahedra respectively. The two types of oxygen are shown in red and light red color. (b) Disordered pyrochlore structure of Ho$_{2}$Zr$_{2}$O$_{7}$. Fig (c) shows the difference between cubes of the fluorite and pyrochlore structures. For fluorite $\alpha$ = 90$^{o}$ (angle between O-A-O) and all the oxygens are equivalent to the central ion. The $\vec{r}_{ij}$ is the unit vector showing displacement of O$^{2-}$ ion from the ideal position. The deviation from fluorite structure can be parameterized in term of $\vec{r}_{ij}$.}
			\vspace{-15pt}
		\end{center}
	\end{figure}
	
X-ray powder diffraction characterized the Ho$_{2}$Zr$_{2}$O$_{7}$ in single phase without any detectable impurity (see Fig. 1). Modelling of the xrd data with the cubic space group is consistent with the disordered pyrochlore structure and gives an excellent fit with Fd$\bar{3}$m space group. Similar to the Dy$_{2}$Zr$_{2}$O$_{7}$ the xrd pattern of Ho$_{2}$Zr$_{2}$O$_{7}$ consists of only the main peaks of pyrochlore structure and the remaining superstructure peaks expected at 2$\theta$ = 14$^{o}$, 27$^{o}$, 36$^{o}$, 42$^{o}$ $\textit{etc.}$ are missing \cite{sheetal2020emergence,mandal2006preparation}. This behavior is consistent with the \textit{r}$_{A}$/\textit{r}$_{B}$ ratio ($\sim$ 1.40) of the compound, which is less than the value at  which a stable pyrochlore structure is expected to evolve (1.48 $\leq$ \textit{r}$_{A}$/\textit{r}$_{B}$ $\leq$ 1.72). Figure 2 shows the crystal structure of stable pyrochlore (a), disordered pyrochlore (b) and a comparison of fluorite and pyrochlore structure (c). 

Unlike the configuration of oxygen atoms in pyrochlore structure the O atoms locally form a perfect cube around both the A and B sites in fluorite structure. The structural change from fluorite to pyrochlore phase which consists the presence of ordered oxygen vacancies, results in the shift in x-coordinate of 48f Oxygen site. The variation in $\alpha$ and $\vec{r}_{ij}$ values can be used as a distortion parameter to study the difference between these structures. Depending on the choice of A and B atoms the crystal symmetry changes and modifies the distortion parameters. It is worth mentioning that the pyrochlore compound La$_{2}$Zr$_{2}$O$_{7}$ with $\alpha$ = 108.65(19) and $\mid$$\vec{r}_{ij}$$\mid$ $>$ 0, shows a change of 12.96$\%$ from fluorite structure ($\alpha$ = 90 and $\mid$$\vec{r}_{ij}$$\mid$ = 0). In Ho$_{2}$Zr$_{2}$O$_{7}$, the obtained $\alpha$ = 91.085(10) and percentage deviation are 0.934$\%$, indicating a small deviation from the fluorite structure. 
	
\begin{figure}[ht]
	\begin{center}
		\includegraphics[width=7cm,height=6cm]{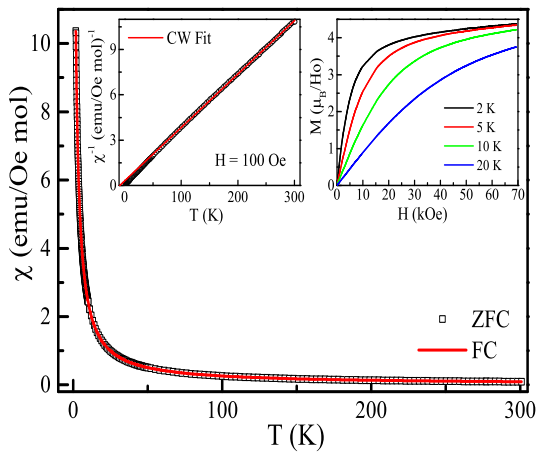}
		\caption{Temperature dependence of dc susceptibility in the range \textit{T} = 1.8 - 300 K. The left inset shows Curie-Weiss fit to the inverse susceptibility data above 30 K. The right inset depicts isothermal magnetization as a function of the applied field showing a saturation moment of $\approx$ 4 $\mu_{B}$/Ho ion at 2 K in addition to a finite slope indicating admixture of higher crystal field (CF) levels.}
		\vspace{-15pt}
	\end{center}
\end{figure}
	
Fitting of the inverse susceptibility data above 80 K (left inset of Fig. 3) by Curie-Weiss law yields an antiferromagnetic Curie-Weiss temperature $\theta_{CW}$ $\sim$ -10.7 (4) K and effective magnetic moment $\mu_{eff}$ $\sim$ 7.6 (05) $\mu_{B}$/Ho. Though the spin ice  pyrochlore systems show crystal field anisotropy, phenomenologicaly the low-temperature (below \textit{T} = 30 K) behavior can again be described by a Curie-Weiss law with the parameters viz. $\theta_{CW}$ $\sim$ -0.5 K and $\mu_{eff}$ $\sim$ 7.2 $\mu_{B}$/Ho. These parameters are similar to that for Dy$_{2}$Zr$_{2}$O$_{7}$.  Here, it is worthy to mention that a ferromagnetic $\theta_{CW}$ $\sim$ +1.9 K was reported for the clean pyrochlore Ho$_{2}$Ti$_{2}$O$_{7}$ system \cite{harris1997geometrical}. The isothermal magnetization as a function of applied field for Ho$_{2}$Zr$_{2}$O$_{7}$ at various temperatures is shown in the right inset of Fig. 3. The saturation magnetization (\textit{M$_{s}$}) at the maximum field (70 kOe) is comparatively less compared to the Ho$_{2}$Ti$_{2}$O$_{7}$ (\textit{M$_{s}$} $\sim$ 5 $\mu_B$/Ho), which is half of the free-ion value and is due to the presence of strong crystal field anisotropy. These data indicate that Ho$_{2}$Zr$_{2}$O$_{7}$ exhibits strong anisotropic behavior like other systems such as Ho$_{2}$Ti$_{2}$O$_{7}$, Dy$_{2}$Ti$_{2}$O$_{7}$ and Dy$_{2}$Zr$_{2}$O$_{7}$ \cite{harris1997geometrical,den2000dipolar,sheetal2020emergence}.
	
	\begin{figure}[ht]
		\begin{center}
			\includegraphics[width=8cm,height=9cm]{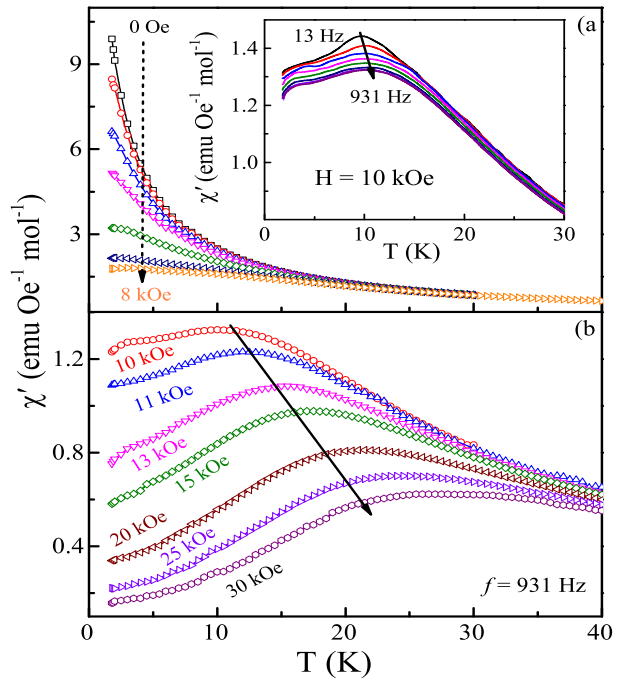}
			\caption{Real part of ac susceptibility data at 931 Hz in the presence of dc magnetic field (\textit{H} $\leq$ 30 kOe). Inset show the frequency dependence of $\chi^{\prime}$ at 10 kOe.}
			\vspace{-15pt}
		\end{center}
	\end{figure}
	
Measurement of the real part of ac susceptibility in the absence of dc magnetic field shows paramagnetic-like behavior down to 1.8 K (Fig. 4a), similar to the behavior of spin ice Ho$_{2}$Ti$_{2}$O$_{7}$, which exhibits spin freezing anomaly at $\sim$ 16 K in the presence of magnetic field \cite{bramwell2001spin}. The large susceptibility value suggests the large spin dynamics down to the lowest measuring temperature. In the presence of a dc magnetic field (\textit{H} $\leq$ 30 kOe). Figure 4b shows the temperature dependence of $\chi^{\prime}$ at various fields at a frequency of 931 Hz. No anomaly appears up to 8 kOe except that of reduction in the magnitude of susceptibility, which indicates slowing down the spin dynamics with the field. A broad hump like feature shows up at \textit{H} = 10 kOe (around 10 K)  and it shifts to higher temperatures with increasing the applied magnetic field. It is worthy of mentioning that Dy$_{2}$Zr$_{2}$O$_{7}$ shows the freezing anomaly in the presence of dc field of 5 kOe. Here, this feature is not seen even up to the field of 8 kOe, which is consistent with the large structural disorder in Ho$_{2}$Zr$_{2}$O$_{7}$ compared to Dy$_{2}$Zr$_{2}$O$_{7}$. It suggests that Ho$_{2}$Zr$_{2}$O$_{7}$ requires higher field to slow down the spin dynamics induced by the large chemical disorder with Zr substitution at the Ti site. It is to note that the structurally ordered Ho$_{2}$Ti$_{2}$O$_{7}$ system shows the thermally activated high-temperature spin freezing behavior in the presence of a magnetic field, which rules out the importance of these features in the formation of the spin ice state at low-temperature \cite{shukla2020robust}. The frequency dependence of freezing anomaly is characterized by calculating the Mydosh parameter (\textit{p} = $\delta$T/\textit{T}$_{f}$ln\textit{f}), which comes out to be 0.18. This large value rules out the spin-glass behavior for which \textit{p} values range between 0.005 to 0.01 and show the unusual spin freezing in these systems \cite{binder1986spin}.\\  
	
	\begin{figure*}[htbp]
		\begin{center}
			\includegraphics[width=18cm,height=6cm]{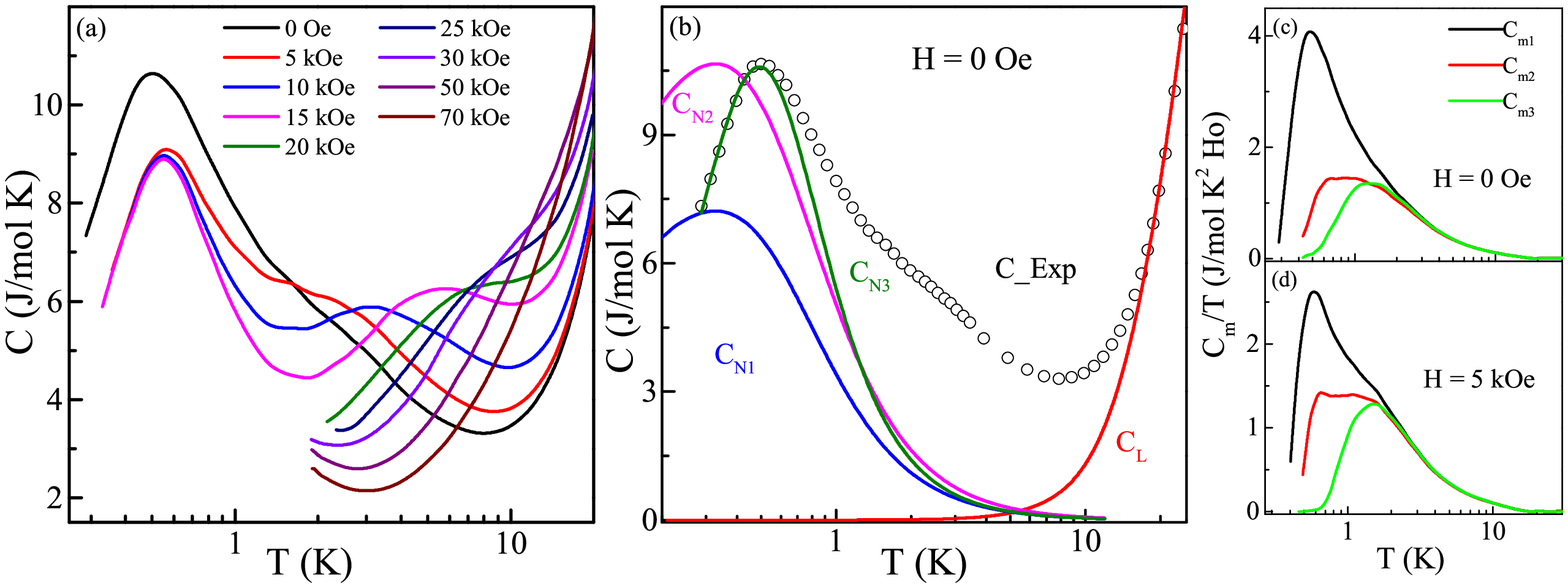}
			\caption{(a) Temperature dependence of heat capacity data at various magnetic fields between 0 - 70 kOe. (b) Heat capacity data analysis of Ho$_{2}$Zr$_{2}$O$_{7}$ at \textit{H} = 10 kOe using various fitting models explained in the main text. (c $\&$ d) shows the temperature dependence of $C_{m}$/T at \textit{H} = 0, 5 kOe. The component $C_{m1}$, $C_{m2}$ and $C_{m3}$ represents the magnetic heat capacity extracted using the different approaches explained in the main text.}
			\vspace{-20pt}
		\end{center}
	\end{figure*} 
	
The specific heat of Ho$_{2}$Zr$_{2}$O$_{7}$ down to 30 mK (shown in Fig. 5) does not show any evidence of phase transition. Instead, the low-temperature heat capacity of the Ho based pyrochlore system shows an increase below $\textit{T}$ = 1 K which is attributed to the anomalously large hyperfine coupling between the nuclear magnetic moment and effective magnetic field (\textit{H}$_{eff}$) at the Ho nucleus, leading to the ambiguities in the exact estimation of the magnetic heat capacity ($C_{m}$). The magnetic heat capacity $C_{m}$ is extracted by subtracting the phonon contributions ($C_{L}$), nuclear contributions ($C_{N}$) and magnetic Schottky anomaly ($C_{sch}$) in case of field data from the total heat capacity. The lattice/phononic contribution to heat capacity was estimated by fitting the high temperature data by using both Einstein and Debye models \cite{sheetal2020emergence}. The value of Debye temperature ($\Theta_D$) and Einstein temperature ($\Theta_E$) were obtained as 122 K and 305 K respectively. For Ho$_{2}$Ti$_{2}$O$_{7}$, the $C_{m}$ was extracted by subtracting the estimated nuclear contribution ($C_{N}$) for the pyrochlore oxide Ho$_{2}$GaSbO$_{7}$ (which shows a peak at $\textit{T}$ = 0.3 K)\cite{bramwell2001spin}. Bl$\ddot{o}$te \textit{et al.} estimated that the $C_{N}$ arises due to the splitting of eight nuclear levels of Holmium (Nuclear isospin \textit{I} = 7/2) with the theoretical maximum of $C_{N}$ $\sim$ 0.9R at $\textit{T}$ = 0.3 K \cite{blote1969heat}. The obtained value of $C_{m}$ for Ho$_{2}$Ti$_{2}$O$_{7}$ using this methodology reveals the same characteristic feature of spin ice as observed for Dy$_{2}$Ti$_{2}$O$_{7}$ \cite{henelius2016refrustration}. Further, the theoretical studies using Monte-Carlo simulation give magnetic entropy values within $\sim$ 2$\%$ of R[ln2-(1/2)ln3/2], which agrees well with the ambit of spin ice materials \cite{den1999comment}. Such a small variance from the Pauling entropy value could be reasonably accounted for any slight deviations in the hyperfine parameters of 4\textit{f} electrons of rare-earth ions (depending upon ionic surrounding, electric field gradient, etc.). Ramon \textit{et al.} extracted the nuclear contribution in Ho$_{2}$Zr$_{2}$O$_{7}$ by subtracting the nuclear contribution for Ho metal \cite{ramon2020geometrically}. However, as the low temperature residual magnetic entropy is one of the important quantities to identify the formation of spin ice state, it is necessary to carefully deal with the fitting models to extract the magnetic contribution. Therefore, we have not restricted ourselves to the approaches used by Bramwell \textit{et al.}, and Ramon \textit{et al.} only for our disordered pyrochlore oxide Ho$_{2}$Zr$_{2}$O$_{7}$ because of its modified crystal field spacing associated with the change in lattice and electronic structure. It was found that the hyperfine interactions of Ho spin lead to a nuclear anomaly at $\sim$ 0.3$\pm$0.02 K \cite{lau2006zero,kumar2016hyperfine,nagata2001specific}. In contrast, the peak position in Ho$_{2}$Zr$_{2}$O$_{7}$ is slightly shifted to a higher temperature of 0.5 K. The subtraction of nuclear part in Ho$_{2}$Zr$_{2}$O$_{7}$ is also an approximation as it overshadows the magnetic contribution below $\textit{T}$ = 1.5 K. We have discussed three possible methods to extract the $C_{m}$ for Ho$_{2}$Zr$_{2}$O$_{7}$ by subtracting the $C_{N}$ (shown in Fig. 5) to illustrate the generic effect of structural symmetry and electronic distribution on the heat capacity. 
	
First, the $C_{m}$ was obtained by subtracting the nuclear hyperfine contribution  $C_{N1}$ used in the previous reports in case of isostructural Ho$_{2}$GaSbO$_{7}$ and Ho$_{2}$Ti$_{2}$O$_{7}$ \cite{blote1969heat,bramwell2001spin}. The obtained $C_{m}$ gives a dynamic ground state crossing the entropy limit of Rln2, expected for the maximum possible states. However, it is hard to rely on this estimation method due to the difference in structural symmetry and effective magnetic moment of Ho$_{2}$Zr$_{2}$O$_{7}$ and Ho$_{2}$GaSbO$_{7}$. Additionally, the observed heat capacity in the systems $\textit{viz.}$ Ho metal, Ho$_{2}$Ti$_{2}$O$_{7}$, Ho$_{2}$GaSbO$_{7}$ pyrochlore, HoCrO$_{3}$ distorted perovskite, shows a  variation of $\approx$ 2-6$\%$ in the peak position and magnitude of $C_{N}$ \cite{krusius1969calorimetric,lau2006zero,kumar2016hyperfine,nagata2001specific}. Therefore in the second case, we have fixed the peak position to 0.3 K and tried to fit the $C_{N}$  of Ho$_{2}$Zr$_{2}$O$_{7}$ (\textit{I} = 7/2) using the equation \cite{lounasmaa1962specific}      
	\begin{eqnarray}
		C_{N} =\frac{R\sum\limits_{i=-I}^{I}\sum\limits_{j=-I}^{I}(W_{i}^{2}-W_{i}W_{j})e^{\frac{-W_{i}-W_{j}}{kT}}}{(kT)^{2}\sum\limits_{i=-I}^{I}\sum\limits_{j=-I}^{I}e^{\frac{-W_{i}-W_{j}} {kT}}} 
		\end{eqnarray}

where R is universal gas constant, \textit{W}$_{i}$/\textit{k} = \textit{a}$^{\prime}$\textit{i} + \textit{P}[\textit{i}$^{2}$ - \textit{I}(\textit{I}+1)/3], \textit{a}$^{\prime}$ (0.299$\pm$0.002 K) is measure of the strength of hyperfine interactions and \textit{P} represents the quadrupole coupling constant which is negligible in case of Holmium. The obtained nuclear hyperfine contribution $C_{N2}$ is shown in Fig. 5b. Remarkably, the additional constraints due to the substitution of Zr for Ti could account for the change in hyperfine interactions, the resulting peak position and magnitude of $C_{N}$. In the third case, we have used the hyperfine interaction model ($C_{N3}$) to fit the raw data without approximating the magnitude or the nuclear peak position. We estimated that the total uncertainty in the extracted data is more than $\approx$ 10$\%$ due to the uncertainty in the subtraction of nuclear spin contribution. Figure 5 (c $\&$ d) shows the extracted $C_{m}$ data at \textit{H} = 0 and 5 kOe using the three different approaches.
	
	\begin{figure}[htbp]
		\begin{center}
			\includegraphics[width=8cm,height=13cm]{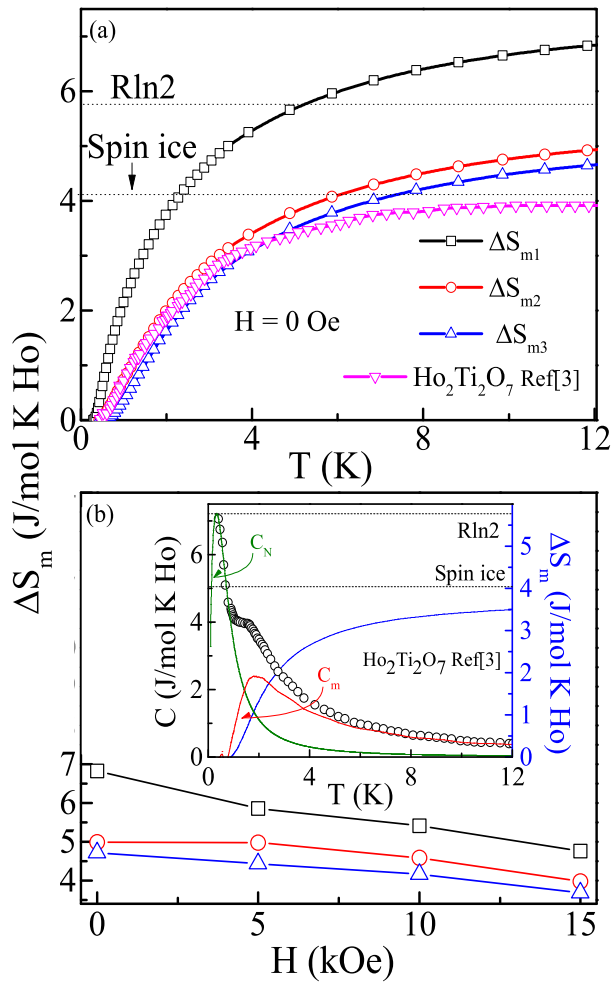}
			\caption{(a) Temperature dependence of extracted magnetic entropy ($\Delta S_{m1}$, $\Delta S_{m2}$ and $\Delta S_{m3}$ extracted using three methodologies) of Ho$_{2}$Zr$_{2}$O$_{7}$ at H = 0 Oe and zero field data of spin ice Ho$_{2}$Ti$_{2}$O$_{7}$ taken from the Ref\cite{bramwell2001spin}. (b) Variation in the magnetic entropy of Ho$_{2}$Zr$_{2}$O$_{7}$ in Ref\cite{bramwell2001spin} as a function of applied field. Inset shows the heat capacity data (symbol) of Ho$_{2}$Ti$_{2}$O$_{7}$ fitted (green) with hyperfine model (Eq. 1), extracted magnetic heat capacity (red) and the magnetic entropy (blue).}
		\end{center}
		\vspace{-20pt}
	\end{figure}
	
The recovered entropy of ($\Delta$S$_{m}$) was calculated by integrating the \textit{C$_{m}$/T} from $\textit{T}$ = 0.3 - 12 K. We have plotted $\Delta$S$_{m}$ of Ho$_{2}$Zr$_{2}$O$_{7}$ obtained at \textit{H} = 0 from the above mentioned three methods along with the $\Delta$S$_{m}$ for Ho$_{2}$Ti$_{2}$O$_{7}$ taken from Ref[3] in Fig. 6a. In the absence of a magnetic field the $\Delta$S$_{m}$ (third case) is comparatively lower than for the disordered state value of Rln2 corresponding to the maximum possible 2$^{N}$ states available to N spins, which indicates that the residual entropy is small compared to the spin ice value as $\textit{T}$ goes to zero. The Ho$_{2}$Ti$_{2}$O$_{7}$ is reported to show Pauling's value of water ice (R/2)ln(3/2) at zero magnetic field \cite{bramwell2001spin}. The integrated entropy for Ho$_{2}$Zr$_{2}$O$_{7}$ is larger than that for Ho$_{2}$Ti$_{2}$O$_{7}$, indicating a significant decrease in the zero-point entropy. The increase in entropy is clearly understood from the chemical alteration of the pyrochlore structure with Zr substitution. The partial replacement of Dy or Ho site in (Dy/Ho)$_{2-x}$Y$_{x}$Ti$_{2}$O$_{7}$ by non-magnetic Y, Lu strongly affect the spin ice state, however, the lattice structure remains intact \cite{lau2006zero,snyder2004quantum}. Figure 6 (b) shows the recovered value of $\Delta$S$_{m}$ at various magnetic fields taken at $\textit{T}$ = 12 K. Application of a magnetic field to the degenerate ground state restored some of the missing entropy, presumably due to the lifted degeneracy of the ground state as the applied field breaks the system’s symmetry. An additional peak appears at $\textit{T}$ $>$ 3 K on switching the magnetic field and shift to higher temperatures with an increase in the magnetic field. This also shows a continuous decrease in magnetic entropy, leading to a non-magnetic ground state at 70 kOe. We have also fitted the heat capacity data of Ho$_{2}$Ti$_{2}$O$_{7}$ taken from Ref [3] using the same hyperfine model (inset of Fig.  6b). Amazingly, the extracted magnetic entropy is significantly lower ($\sim$ 12$\%$) than the expected value for the spin ice systems. Thus the approach of using the nuclear contribution of Ho$_{2}$GaSbO$_{7}$ by Bl$\ddot{o}$te \textit{et al.} is not a good choice for the estimation of magnetic heat capacity of  Ho$_{2}$Ti$_{2}$O$_{7}$ compound.\\
	
In conclusion, Ho$_{2}$Zr$_{2}$O$_{7}$ with its disordered pyrochlore structure shows the presence of spin freezing at \textit{T} $\sim$ 10 K on the application of magnetic field. Unlike the spin ice systems, the heat capacity and susceptibility indicate large spin dynamics at the lowest measuring temperature and the absence of Pauling residual entropy. These studies show that the low-temperature magnetic state of pyrochlores systems can be significantly altered by the frustration and the structural disorder.  We have further analyzed the low-temperature heat capacity and discussed the error arising from the estimation of nuclear contribution in Holmium based system. We observed that the use of the hyperfine model (Eq. 1) for the estimation of nuclear heat capacity leads to the significant correction in the residual spin ice entropy of the Ho$_{2}$Ti$_{2}$O$_{7}$. The disordered pyrochlore zirconates provide an excellent family of materials for further investigation, for the role of induced disorder on the spin dynamics and the creation/propagation of monopole-antimonopole excitations in the frustrated systems. The low-temperature neutron scattering technique would be important for the identification of true magnetic ground state in Ho$_{2}$Zr$_{2}$O$_{7}$.\\

\textbf{Acknowledgement:} We thank AMRC, Indian Institute of Technology Mandi for the experimental facility. Sheetal acknowledged IIT Mandi and MHRD India for the HTRA fellowship. We acknowledge financial support by BMBF via the project SpinFun (13XP5088) and by Deutsche Forschungsgemeinschaft (DFG) under Germany’s Excellence Strategy EXC2181/1-390900948 (the Heidelberg STRUCTURES Excellence Cluster) and through project KL 1824/13-1.   
	
	\bibliography{Ref}
	
\end{document}